\documentclass[aps,reprint,prd,showpacs,amsmath,amssymb,nofootinbib]{revtex4-1}
\usepackage{graphicx}
\usepackage{color}
\usepackage{times}
\usepackage{inputenc}
\usepackage{bm}
\usepackage{multirow}
\usepackage{float}
\usepackage{url}
\usepackage{natbib}
\usepackage{tensor}
\usepackage{xcolor}
\usepackage[colorlinks=true,citecolor=blue,urlcolor=blue,linkcolor=red]{hyperref}
\usepackage[normalem]{ulem}
\usepackage{appendix}

\def\msun{\,M_{\odot}}
\def\fm3{\;\text{fm}^{-3}}
\def\mev{\;\text{MeV}}

\begin{document}

	\title{On the survival of strong nuggets in the early Universe}
	
	\author{Haoyang Qi$^{1,2}$}
	\email{qhy021@pku.edu.cn}
	\author{Wen-Li Yuan$^{1,2}$}
	\email{wlyuan@pku.edu.cn}
	\author{Yudong Luo$^{2,3}$}
	\email{yudong.luo@pku.edu.cn}
	\author{Chao Chen$^{4}$}
	
	\author{Shi Pi$^{5,6,7}$}
	
	\author{Renxin Xu$^{1,2}$}
	\email{r.x.xu@pku.edu.cn}
	\affiliation{$^1$State Key Laboratory of Nuclear Physics and Technology, Peking University, Beijing 100871, China;
	}
	\affiliation{$^2$Department of Astronomy, School of Physics, Peking University, Beijing 100871, China}
	\affiliation{$^3$Kavli Institute for Astronomy and Astrophysics, Peking University, Beijing 100871, China}
	\affiliation{$^4$School of Science, Jiangsu University of Science and Technology, Zhenjiang 212100, China}
	\affiliation{$^5$Institute of Theoretical Physics, Chinese Academy of Sciences, Beijing 100190, China}
	\affiliation{$^6$ Center for High Energy Physics, Peking University, Beijing 100871, China }
	\affiliation{$^7$ Kavli Institute for the Physics and Mathematics of the Universe (WPI), The University of Tokyo, Kashiwa, Chiba 277-8583, Japan}
	
	\begin{abstract}
        Strong nuggets with a baryon number of $A\sim 10^{10-30}$ could be able to survive from the cosmic separation of the QCD phases, provided the transition from strange quark matter to strangeon matter is accounted for, thereby evading evaporation in the early Universe. Such strangeon nuggets may serve as a dark matter candidate within particle standard model. We formulate the corresponding phase transition of cosmic strange matter, establishing a parameter space which reasonably accommodates observational constraints on the dark-to-luminous matter ratio and the mass-radius relation, as well as tidal deformability of compact objects.

	\end{abstract}
	\maketitle
	
	\section{Introduction}\label{sec:intro}
	
	Of the four fundamental interactions, the strong one presents experimentally realistic challenges, i.e., understanding behaviours at the low-energy scale.
	Quantum chromodynamics (QCD) as the underlying physics is successful in perturbative high-energy regime, but its non-perturbative calculations are relevant to one of the Millennium Prize Problems, particularly in the case of a high baryon chemical potential~\cite{klevansky_nambu---jona-lasinio_1992,buballa_njlmodel_2005,2008RvMP...80.1455A,2017EPJWC.14104001L,2018RPPh...81e6902B}.
	Both atomic nuclei and Landau’s ``gigantic nucleus''~\cite{landau_theory_1932} fall within this scope, and the latter, being the most challenging problem in multimessenger astronomy (especially following the discovery of gravitational waves from binary neutron star (NS) mergers~\cite{ligo_observation_2017,ligo_gw170817_2018,abbott_properties_2019}), is the focus of the era: the equation of state (EoS) of supranuclear matter~\cite{2008RvMP...80.1455A,2017EPJWC.14104001L,2018RPPh...81e6902B,2025SCPMA..6819503L}.
	Additionally, understanding the nature of strong-interaction matter (or simply ``strong matter'') is also significant for explaining the QCD phase transition in the early Universe \cite{witten_cosmic_1984}, and the surviving strong nuggets are potential candidates for dark matter (DM) in the context of the standard model.
	This paper addresses these two issues.
 
    The EoS of dense matter, which determines the internal structure, stability, and observable properties—such as mass, radius, and tidal deformability—remains largely uncertain~\cite{2008RvMP...80.1455A,2017EPJWC.14104001L,2018RPPh...81e6902B,2025SCPMA..6819503L}.
    As a result, all compact stars could be entirely composed of deconfined quarks, forming quark stars self-bound on surface~\cite{1998PhLB..438..123D,1989PhLB..229..112C,1999PhLB..457..261B,1999PhRvC..61a5201P,2005PhRvC..72a5204W,2010MNRAS.402.2715L,2018PhRvD..97h3015Z,holdom_quark_2018,2018PhRvD..98h3013L,2019PhRvD..99j3017X,2000PhRvC..62b5801P,2021EPJC...81..612B,2021ChPhC..45e5104X,2022PhRvD.105l3004Y,2024PhRvD.110j3012Z,2024FrASS..1109463Z}. Nonperturbative QCD effects may lead to the localization of quarks into clusters. Such clustered states, originally referred to as strange quark clusters, are now commonly known as strangeons, which can be regarded as nucleon-like bound states~\cite{xu_solid_2003,2025arXiv251101146X}. Strangeon matter is characterized by an intrinsically stiff EoS~\cite{2006MNRAS.373L..85X,2008MNRAS.384.1034P,lai_lennardjones_2009,2012MNRAS.424.2994L,2014MNRAS.443.2705Z,2016ChPhC..40i5102L,2018RAA....18...24L,2019EPJA...55...60L,lai_merging_2021,2022MNRAS.509.2758G,2022IJMPE..3150037M,2023PhRvD.108f3002Z,2023PhRvD.108l3031Z}, and was predicted to support massive pulsars with masses exceeding $2\msun$ even before the discovery of the first precisely measured massive pulsar PSR J1614–2230~\cite{lai_lennardjones_2009,2010Natur.467.1081D}. The potential existence of ultralow-mass compact stars has also sparked interest in speculating about strange stars as the potential physical nature of pulsar-like compact objects~\cite{2015ApJ...798...56L,2022NatAs...6.1444D}.

    Beyond employing astronomical observations to deepen our understanding of cold strong matter inside compact objects, it is also fascinating to explore its consequences during the cosmic QCD phase transition at finite temperatures~\cite{chatrchyan_gravitational_2026}. 
    Although lattice gauge theory at low net baryon density indicates a rapid crossover from the quark-gluon plasma (QGP) to the hadronic phase~\cite{karsch_lattice_2000,rajagopal_condensed_2001,fukushima_phase_2010}, the specific structure of the QCD phase diagram remains far from fully understood. The distinctive and potentially observable imprints of a first-order phase transition in the early Universe have attracted considerable interest~\cite[e.g.,][]{gao_cosmology_2022,chatrchyan_gravitational_2026}. 
    
    In this context, dense strange quark nuggets may be generated during a first-order QCD phase transition in the early Universe~\cite{witten_cosmic_1984}. As the Universe cools down, it separates into a high-temperature QGP phase and a low-temperature hadronic gas phase.
    The hadronic phase has a low baryon number density since baryons are non-relativistic particles and obey the Maxwell-Boltzmann distribution due to their large mass. Therefore,
    when high- and low-temperature phases reach thermal equilibrium, most of the net baryon number is concentrated in the high-temperature regions. As cosmic expansion continues,
    these high-temperature domains lose energy and their boundaries gradually contract. Witten hypothesized that, during this contraction process, the baryon number is effectively trapped within the boundaries~\cite{witten_cosmic_1984}. Consequently, the shrinking QGP regions approach nuclear saturation density, ultimately culminating in the formation of dense strange quark nuggets.

    Several works investigated the evaporation of strange quark nuggets or strangeon nuggets at high temperature~\cite{alcock_evaporation_1985,li_revisiting_2015,pietri_merger_2019,lai_merging_2021,bucciantini_formation_2022,miao_equation_2025}. In the hot and dilute environment of the early Universe, Alcock and Farhi~\cite{alcock_evaporation_1985} displayed that strange quark nuggets with baryon numbers below approximately $10^{52}$ would not survive. 
    Due to particle horizon constraints during the cosmic QCD phase transition, the formation of strange quark nuggets with baryon numbers greater than $10^{52}$ is highly suppressed, as they require density fluctuations $10^3$ times greater than the cosmic average~\cite{alcock_evaporation_1985}.  However, a strangeon nugget evaporates much slower than a strange quark nugget because of the additional barriers of strangeon matter~\cite{lai_merging_2021,xu_solution_2014,wang_optical_2017,li_polar_2026}. 
    When nucleons evaporate from a strange quark nugget, they are assembled from up quarks and down quarks quickly ($\sim 10^{-23}$\,s for typical strong interaction~\cite{bodmer_collapsed_1971,witten_cosmic_1984,alcock_evaporation_1985,kamal_particle_2014}). By contrast, if nucleons evaporate from the strangeon nugget surface, either the strangeons emit and then decay into nucleons, or the nucleons form within the strangeon matter and then emit:
    for the former, the emission of strangeons needs to overcome a strong barrier and is thus suppressed~\cite{lai_merging_2021,madsen_rate_1993}, for the latter, a much slower weak interaction is necessary to convert a strangeon into nucleon(s)~\cite{xu_solution_2014,wang_optical_2017,li_polar_2026}.\footnote{%
    In view of a thermal equilibrium for calculating the emissivity of nucleons from strong nuggets, let us discuss its reverse process of the nucleon evaporation, with explanation of an astrophysical example.
    Accreted nucleons (e.g., protons) with a kinematic energy $\sim 10^2$ MeV would be easily absorbed by a bare strange quark star. However, because of the strangeness barrier and the hard core between nucleon and strangeon, those nucleons rebound from a strangeon matter surface~\cite{xu_solution_2014}, that would be the cause of a peculiar atmosphere~\cite{wang_optical_2017} and a polar mound~\cite{li_polar_2026} in astrophysics.
    }
    These result that the nucleon evaporation rate of strangeon nuggets is $<10^{-13}$ times that of strange quark nuggets, and consequently that a large number of strangeon nuggets with $A<10^{49}$ can survive. This makes them candidates for DM. 
    
    Similar to strange quark matter, strangeon matter contains nearly equal proportions of up, down, and strange quarks. Its charge-to-mass ratio is approximately $10^{-4}$ or lower, resulting in an extremely weak electromagnetic signal that can hardly be detected over significant distances. This characteristic positions strangeon matter as a potential DM candidate, especially since the cross-section of such strong nuggets is expected to comply with existing observational constraints \cite{carlson_selfinteracting_1992,popolo_neutron_2020,tseng_phenomenology_2024,nadler_cozmic_2025} (for a comprehensive review, see \cite{cirelli_dark_2024}).

    This paper explores how strangeon nuggets survived in the early Universe by undergoing a transition from strange quark nuggets. 
    It is organized as follows. Section~\ref{sec:formalism} 
    introduces the EoSs 
    of strange quark matter and strangeon matter at finite temperature and density. In Section~\ref{sec:evaporation},
    we calculate the evaporation of strange quark nuggets and explore the parameter space of the strangeon EoS that allow strangeon nuggets to survival as a DM candidate. 
    We make the conclusion in Section~\ref{sec:summary}.

	\section{formalism}\label{sec:formalism}

	\subsection{Strange quark matter at finite temperature}
	We employ the MIT bag model, which captures the confinement property of QCD, to describe strange quark matter. In this framework, quark interactions are effectively encoded in the perturbative QCD vacuum energy density $B$, defined as the excess of the energy density of the QCD vacuum inside the bag over that of the physical vacuum outside the bag. The grand canonical potential density $\Omega$ of strange quark matter~\cite{chodos_new_1974,chodos_baryon_1974,johnson_mit_1975,fisher_theory_1967} is given by:
	\begin{equation}
			\begin{aligned}
				\Omega &= -6\sum_{i=u,d,s}\int \frac{{\rm d}^3 k}{(2\pi)^3} \frac{k^2}{3E_i} \left( f_{i} + \bar{f_{i}} \right)+B~,
				\label{eq:MITOmega}
			\end{aligned}
	\end{equation} 
	in which
	\begin{equation}
		\begin{aligned}
			f_{i}(k,\mu_i,T) &= \frac{1}{1+\exp \left(\frac{E_i-\mu_i}{T}\right)},\\
			\bar{f_{i}}(k,\mu_i,T) &= \frac{1}{1+\exp \left(\frac{E_i+\mu_i}{T}\right)},\\
		\end{aligned}
	\end{equation}
	with $E_i=\sqrt{k^2+m_i^2}$.
	From the grand canonical potential density in Eq.~(\ref{eq:MITOmega}), all thermodynamic quantities of interest can be calculated using standard thermodynamic relations~\cite{alcock_strange_1986}. The pressure, the quark number densities, and the energy density of strange quark matter are
	\begin{equation}
		\begin{split}
			p &=-\Omega = 6\sum_{i=u,d,s}\int \frac{{\rm d}^3 k}{(2\pi)^3} \frac{k^2}{3E_i} \left( f_{i} + \bar{f_{i}} \right)-B~,\\
			n_i &=  
			- \frac{\partial \Omega}{\partial \mu_i}
			= 6\int \frac{{\rm d}^3 k}{(2\pi)^3} \left( f_{i} - \bar{f_{i}} \right)~, (i=u,d,s)~,\\
			\varepsilon &= \Omega - Ts  
			+ \sum_{i=u,d,s}\mu_i n_i \\
			&= 6\int \frac{{\rm d}^3 k}{(2\pi)^3} E_i \left( f_{i} + \bar{f_{i}} \right)+B~.
		\end{split}
	\end{equation}

	In $\beta$-equilibrium matter, the relations between the chemical potentials of different particles are given by
		\begin{equation}
			\begin{split}
				\mu_s = \mu_d=&\mu_u + \mu_e\ ,\\
				\mu_{\mu}=&\mu_{e}~.
			\end{split}
		\end{equation}
	The charge-neutrality of the strange quark matter and the baryon number conservation should also be required
		\begin{equation}
			\begin{split}
				\frac{2}{3}n_u - \frac{1}{3}n_d - \frac{1}{3}n_s - n_e - n_\mu = 0~,\\
				n_{\rm B} = \frac{1}{3}(n_u +n_d +n_s)~.\\
			\end{split}
		\end{equation}
	
	Then the total energy density and total pressure of the dense matter are obtained by including both the contribution of quarks and leptons
		\begin{equation}
			\begin{aligned}
				\mathcal{E} &= \varepsilon_{\text {quark }} +\varepsilon_{\text {lep }}, \quad 
				P &= p_{\text {quark }}+ p_{\text {lep }}.
			\end{aligned}
		\end{equation}
    
	Given the down quark chemical potential $\mu_d$,
	assuming that neutrinos are free, we can calculate all the thermodynamic quantities and obtain the EoS.
	The current quark masses $(m_u = 2.2\,{\rm MeV}$, $m_d = 4.7\,{\rm MeV}$, $m_s = 94\,{\rm MeV})$  and lepton masses $(m_e = 0.511\,{\rm MeV}$, $m_\mu = 105.7\,{\rm MeV})$ we used here are from the latest Review of Particle Physics~\cite{navas_review_2024}.
	
	\vspace{6pt}
	
	\subsection{Strangeon matter at finite temperature}\label{subsec:Strangeon}
    
	A strangeon is a color-confined state that interacts predominantly through residual strong interactions with surrounding strangeons, in close analogy to the interactions between nucleons~\cite{xu_solid_2003,lai_lennardjones_2009,kamal_particle_2014}.
	Therefore, the double Yukawa potential~\cite{yukawa_interaction_1935,faessler_nuclear_1997,lai_hcluster_2013} is employed here to describe the interaction between two strangeons, 
    \begin{equation}\label{eq:YukawaPotential}
		\begin{split}
			V(r) = g_1 \frac{e^{-m_1 r}}{r}
			- g_2 \frac{e^{-m_2 r}}{r}~,
		\end{split}
	\end{equation}
	from which we derive the thermodynamical quantities of strangeon matter.

	Imposing the conditions 
	\begin{equation}
		V(r_0) =  
		-u_0~,~~~~ \frac{dV}{dr}\bigg\vert_{r_0}=0~,
	\end{equation}
	the coupling constants $g_1$ and $g_2$ can be reparametrized as follows:
	
    \begin{equation}
		\begin{split}
			g_1 &= \frac{u_0}{m_1-m_2}e^{m_1 r_0}(1+m_2 r_0)~, \\
			g_2 &= \frac{u_0}{m_1-m_2}e^{m_2 r_0}(1+m_1 r_0)~, 
		\end{split}
	\end{equation}
	where $r_0$ denotes the equilibrium separation between two strangeons, at which the attractive and repulsive forces exactly equilibrium and the potential reaches its minimum value $-u_0$.
    Since $r_0$ and $u_0$ describes the equilibrium state, their value should vary against temperature. To incorporate such effects, we use $T_0$ to quantify the shallowing of the potential well and the lattice expansion as the temperature increases. We introduce an explicit temperature dependence for the parameters $u_0$ and $r_0$,
	\begin{equation}\label{eq:T0}
		\begin{split}
			u_{0}&= 
			u_{0}^{(0)}\exp{(-T/T_0)}~,\\
    			r_0 &= r_0^{(0)}\exp{(T/T_0)}~, 
		\end{split}
	\end{equation}
	in which $u_{0}^{(0)}$ and $r_0^{(0)}$ denote the values of $u_0$ and $r_0$ at zero temperature. 
    In the present study, $T_0$ is expected to be of the same order of magnitude as the QCD deconfinement transition temperature, around $100\mev$~\cite{karsch_lattice_2000,rajagopal_condensed_2001,fukushima_phase_2010}.

    At sufficiently low temperatures, strangeons may form a crystalline lattice as a result of their strong mutual interactions~\cite{lai_lennardjones_2009}. In this case, the potential energy of a single strangeon can be evaluated by summing its interactions with all other strangeons located on a simple cubic lattice relative to the origin. The total potential energy density is then obtained by multiplying by the strangeon number density $n$ and dividing by a factor of two to avoid double counting,
    \begin{equation}
		\varepsilon_{\rm p}=\frac{n}{2}\sum_{\substack{(N_1,N_2,N_3) \\ \in \mathbb{Z}^3\backslash(0,0,0)}}V\left(n^{-1/3}\sqrt{N_1^2+N_2^2+N_3^2}\right)~.
	\end{equation}
    Here, $n^{-1/3}$ corresponds to the lattice constant, i.e., the distance between nearest-neighbor strangeons.

	With increasing temperature, strangeon matter may undergo melting, analogous to the dissociation of hadrons into a QGP at high temperatures~\cite{gross_qcd_1981,pasechnik_phenomenological_2017,andronic_decoding_2018}. However, due to the rapid decay of the double Yukawa potential with distance, the dominant contribution to the potential energy arises from nearest-neighbor interactions, which do not differ significantly between the solid and liquid configurations. Therefore, the lattice-based treatment mentioned above remains a good approximation over the entire temperature range we considered in this work.

    Owing to the large mass of strangeons, their de Broglie wavelengths are much smaller than the inter-strangeon spacing, allowing them to be treated as classical particles. Consequently, the thermal motion of strangeons at finite temperature can be well described by the Maxwell-Boltzmann distribution, from which the entropy density and the thermal kinetic energy are derived as follows:
    \begin{equation}
        \begin{aligned}
            s &= \frac{5}{2} n
            + n \ln\!\left[\left(m T/2\pi\right)^{3/2} n^{-1}\right], \\
            \varepsilon_{\rm k} &= \frac{3}{2} n T~.
        \end{aligned}
    \end{equation}
    Then the total energy density therefore consists of the potential energy, rest energy, and thermal kinetic energy,
    \begin{equation}
    \begin{aligned}
		\varepsilon = \varepsilon_{\rm p} + m n + \varepsilon_{\rm k}\ ,
        \end{aligned}
	\end{equation}   
    where $m$ is the rest mass of a strangeon. 
    Employing the standard thermodynamic relations, the pressure and baryon chemical potential read,	
	\begin{equation}\label{eq:p}
		\begin{split}
			p &= -T n^2 \frac{{\rm d}(s/n)}{{\rm d}n} + n^2 \frac{{\rm d}(\varepsilon/n)}{{\rm d}n}~,\\
			\mu_{\rm B} &= \frac{\varepsilon + p - Ts}{n_{\rm B}}~.
		\end{split}
	\end{equation}

    Aside from $T_0$ and $m$, the strangeon matter EoS contains four free parameters, $u_0^{(0)}$, $r_0^{(0)}$,  $m_1$ and $m_2$. Nucleon-nucleon scattering data indicate that the inter-nucleon potential well lies in the range $\sim 50 -120 \mev$ for the ${}^{1}\text{S}_{0}$ (spin-singlet and S-wave) channel~\cite{1994PhRvC..49.2950S,1995PhRvC..51...38W,2001PhRvC..63b4001M}. Since the strong interactions are not sensitive to the flavor of quarks, this work sets the range of $u_0^{(0)}$ to $0\sim400\,\mev$.  The equilibrium distance, $r_0^{(0)}$, is determined by ensuring that  the density of strangeon matter exceeds the nuclear saturation density $n_0$. 
    The masses $m_1$ and $m_2$ reflects the force range of the interactions, with $r_{\rm range}=1/m$. 
    The values of $m_1$ and $m_2$ are chosen in the range of $60\sim1200\,\mev$, with the constraint $m_1>m_2$, reflecting the expectation that the repulsive interaction has a shorter range than the attractive one. Regarding the input parameter of the single strangeon mass, $m$. The requirement for a strangeon to be a color singlet implies that it must consist of an integer multiple of three quarks. Recent Bayesian inference studies of strangeon matter, incorporating NS mass and radius measurements together with the gravitational-wave event GW170817, indicate that a strangeon favors a stable bound state with the quark number in a strangeon $N_{\rm q} = 18$~\cite{yuan_bayesian_2025}. We therefore adopt the strangeon mass as $m = N_{\rm q}m_{\rm q} = 18 \times 310\,{\rm MeV} = 5580\,{\rm MeV}$, where $m_{\rm q}=310\,{\rm MeV}$ is the effective quark mass.
    The value is one third of the energy per baryon of $^{56}$Fe nuclei. To satisfy the assumption that stangeon matter is more stable than $^{56}$Fe at zero temperature and pressure, the effective quark mass cannot be much larger than $310\mev$. A simple estimation suggests that taking $m_{\rm q}$ from $260\mev$ to $320\mev$ or $N_{\rm q}$ from 15 to 21 does not have a significant impact on the conclusion.
 

	\section{Evaporation and QCD phase transition}\label{sec:evaporation}
	Strange quark nuggets may be produced as the Universe cools down through a first-order QCD phase transition~\cite{witten_cosmic_1984}. As mentioned in Sec.~\ref{sec:intro}, a subsequent first-order phase transition--the conversion of strange quark nuggets into strangeon nuggets--can substantially suppress the evaporation rate, enabling the strangeon nuggets to survive and rendering them viable DM candidates~\cite{alcock_evaporation_1985,madsen_rate_1993,xu_solution_2014,wang_optical_2017,li_polar_2026}. In this section, we analyze the evolution of the strange quark nuggets and determine the characteristic temperature at which this transition occurs.
    
	The cosmic expansion is described by the Friedmann equation \cite{ryden_introduction_2016}. Combine it with the EoS in the radiation-dominated Universe, i.e., $\overline{p}=\overline{\varepsilon}/3$, the average energy density of the Universe $\overline{\varepsilon}$ and cosmic temperature $T$ evolution are then given by
	\begin{equation}
		\begin{split}
			\overline{\varepsilon}&\propto a^{-4} \propto t^{-2}~, \\
			T &\propto \overline{\varepsilon}~^{1/4} \propto t^{-1/2}~,
		\end{split}
	\end{equation}
	where $a$ is the scalar factor.
	QCD deconfinement phase transition occurs at $T\simeq 100$\ MeV~\cite{karsch_lattice_2000,rajagopal_condensed_2001,fukushima_phase_2010}, at this epoch, the average baryon density of the Universe is extremely low, approximately $10^{-9} n_0$, and the environment primarily consists of photons, neutrinos, electrons, and their antiparticles, therefore the evaporation of strange quark nuggets can follow the formulae from Alcock and Farhi's work~\cite{alcock_evaporation_1985} (see Appendix~\ref{appendix: thermo} for the short review).
    Some existing works have calculated the evaporation of strange quark nuggets or strangeon nuggets~\cite{li_revisiting_2015,lai_merging_2021,bucciantini_formation_2022,miao_equation_2025}. They either considered a more refined quark mass density-dependent EoS, or took into account the relatively high-density environment (which can reach up to $10^{-3}n_0$) in the ejecta of merging binary NSs. In this work, we use a simple MIT bag model to investigate the possibility that strange quark nuggets could stop evaporating and survive. The simple quark matter EoS does not affect the conclusion qualitatively or semi-quantitatively. Additionally, The issue of evaporation saturation caused by a high-density nuclear gas environment does not need to be considered in the early Universe.

	   \begin{figure}
		\includegraphics[width=0.48\textwidth]{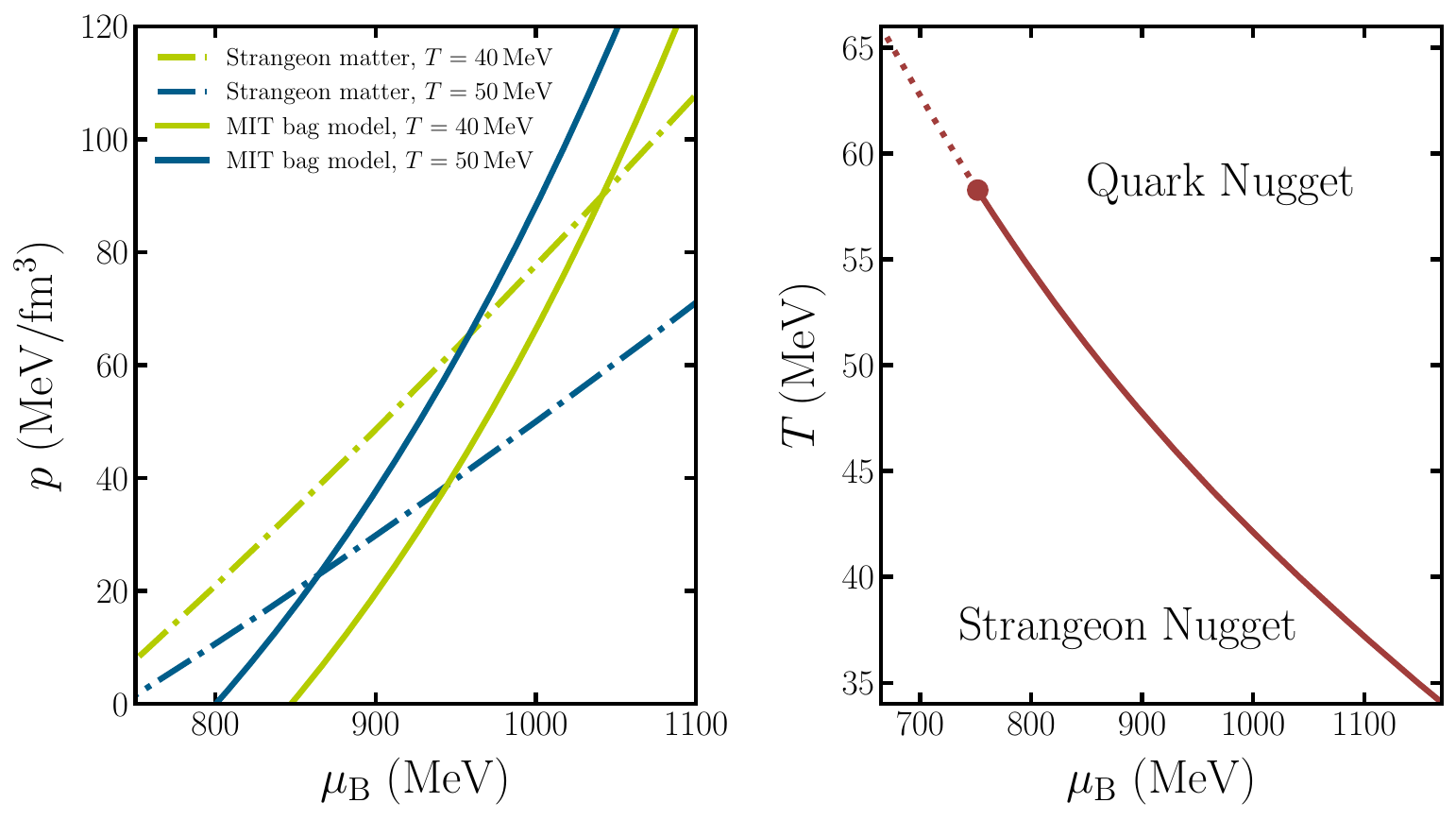}
        \caption{Left panel:  Pressure $p$ as a function of the baryon chemical potential $\mu_{\rm B}$ for strange quark matter and strangeon matter. The MIT bag model calculations for strange quark matter with $(B^{1/4}=159~\mathrm{MeV}, T=40~\mathrm{MeV})$ and $(B^{1/4}=159~\mathrm{MeV}, T=50~\mathrm{MeV})$ are shown as the solid-green and solid-blue curves, respectively. The dot-dashed curves represent the results for strangeon matter with $(m_1=710~\mathrm{MeV}, m_2=239~\mathrm{MeV}, u_{0}^{(0)}=303~\mathrm{MeV}$, $r_0^{(0)}=2.06~\mathrm{fm})$ at $T=40~\mathrm{MeV}$ and $T=50~\mathrm{MeV}$. 
        Right panel: The strange quark-strangeon phase diagram in the $T–\mu_B$ plane.
        }\label{fig:pmuBTmuB}
	\end{figure}

     \begin{figure}
		\includegraphics[width=0.48\textwidth]{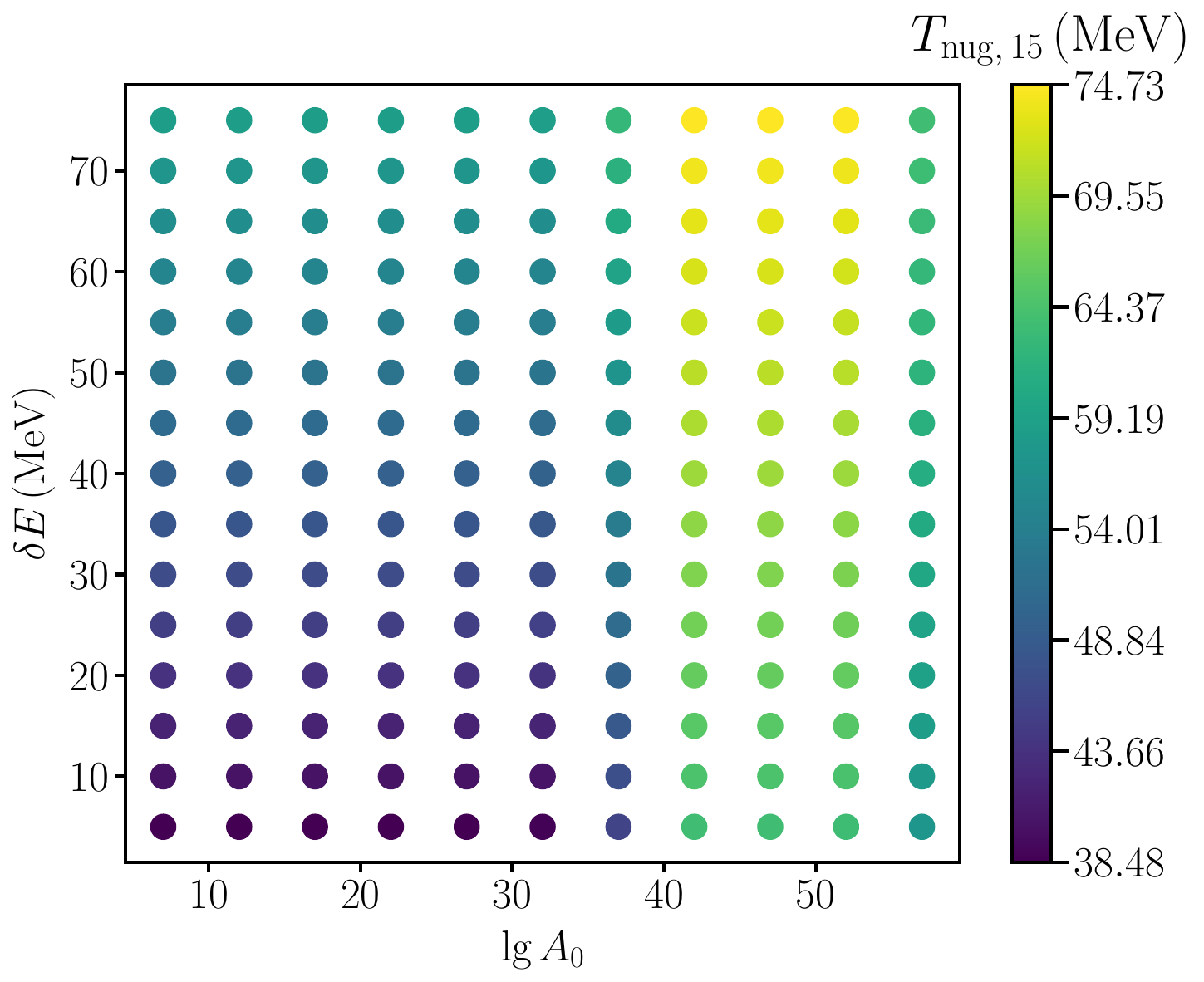}
        \caption{The strange quark nugget temperature at which it loses 15\% of its baryon number due to evaporation (we define it as $T_{\rm nug,\,15}$). In the calculation, we assume the strange quark nugget forming temperature $T_{\rm u0} = 100\,{\rm MeV}$, with initial baryon number $A_0$ ranged from $10^7$ to $10^{57}$ and binding energy per baryon $\delta E$ from 0 to $75\mev$.
        }\label{fig:T15}
	\end{figure}
    
    \begin{figure*}		
\includegraphics[width=0.48\textwidth]{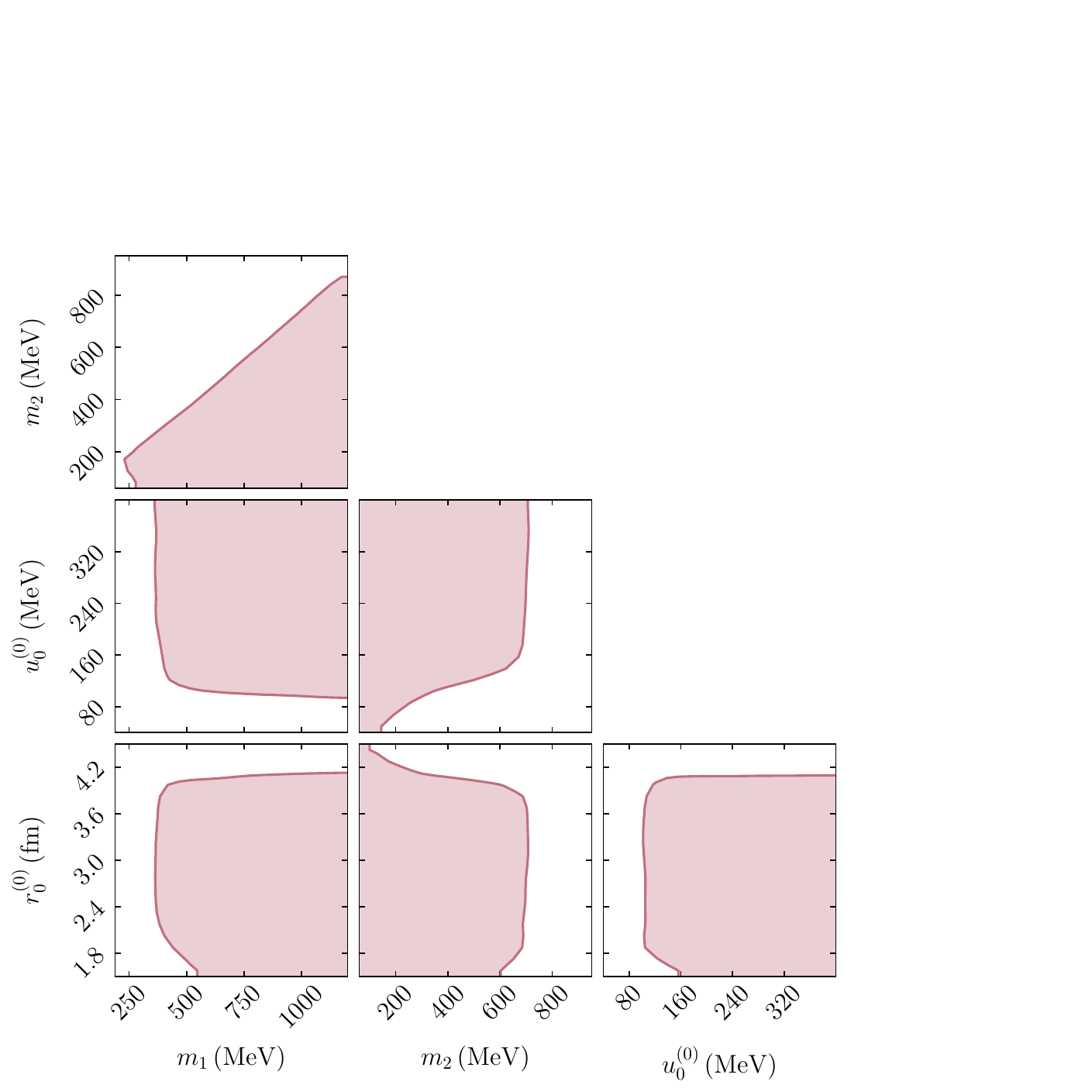}
\includegraphics[width=0.48\textwidth]{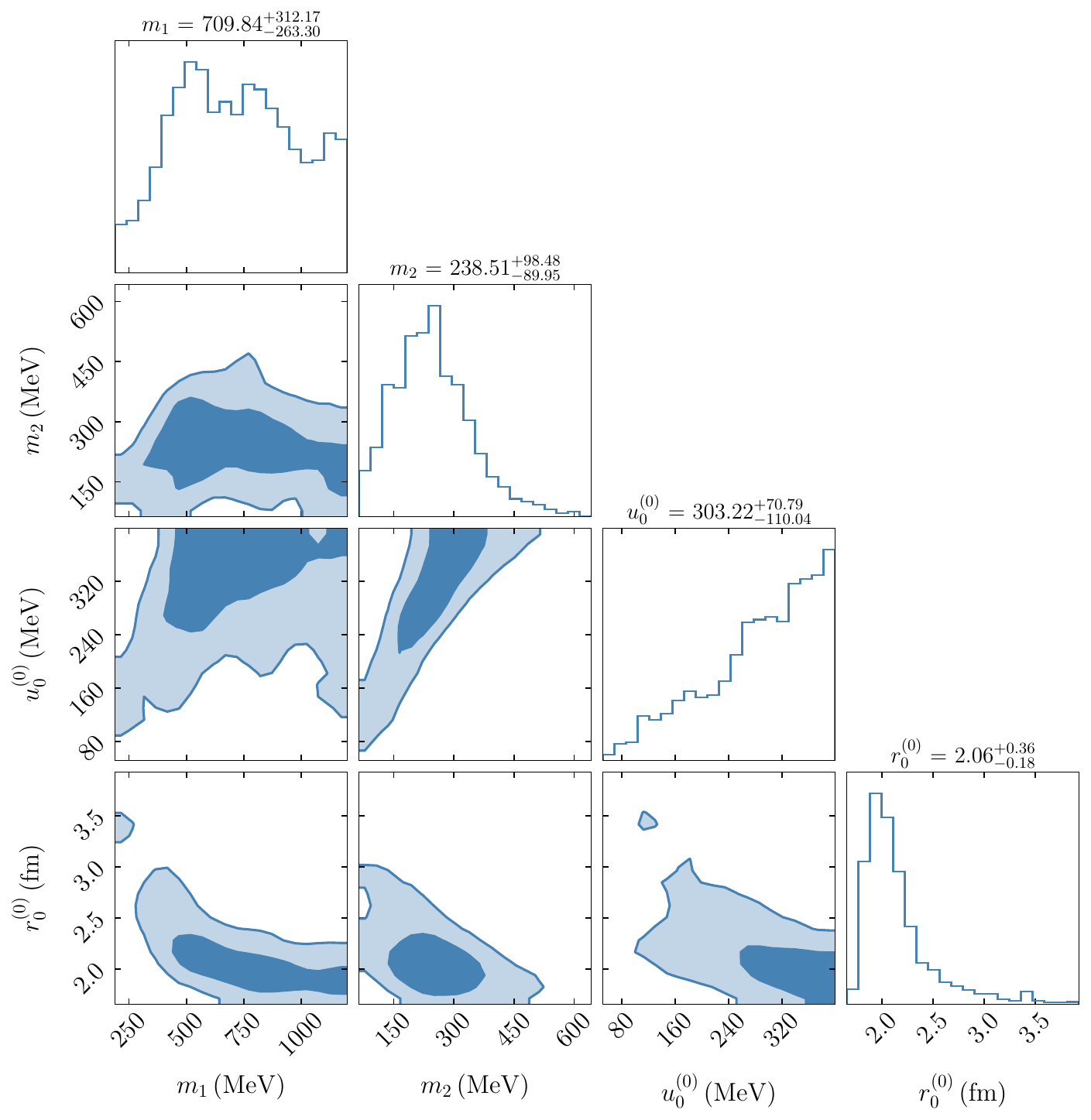}
	    \caption{
        Left panel: the parameter space of strangeon EoS in which the corresponding phase transition temperature cause strange quark nugeets to undergo phase transition to strangeon nuggets at a certain $A_0$ and $\delta E$, retaining 85\% of its baryon number. The red area contains 90\% of the sampled dots. In the calculation, $T_0$ in Eq.~(\ref{eq:T0}) is set to 100\,MeV, at the same level as the QCD phase transition temperature~\cite{karsch_lattice_2000,rajagopal_condensed_2001,fukushima_phase_2010}. Right panel: the parameter space of strangeon EoS obtained by a Bayesian inference using the mass-radius measurements from NICER, such as PSR J0030+0451~\cite{riley_nicer_2019,vinciguerra_updated_2024,miller_psr_2019}, PSR J0437-4715~\cite{choudhury_nicer_2024,rutherford_constraining_2024}, PSR J0740+6620~\cite{miller_radius_2021,riley_nicer_2021}, and the tidal deformability measurement of LIGO/Virgo from GW170817~\cite{ligo_observation_2017,ligo_gw170817_2018,abbott_properties_2019}, within the allowed parameter space in the left panel. The contour confidence levels are 50\% and 90\%.
       }\label{fig:params}
	\end{figure*}

Considering the evaporation of strange quark nuggets in the high-temperature and low-density environment of the early Universe, the evaporation rate calculated from Eq.~(\ref{eq:dAdt}) is about $10^{58}{\rm s^{-1}}(A/10^{52})^{2/3}$ for baryon number $A>10^{40}$, binding energy per baryon $\delta E=40\mev$, and temperature of the Universe $T_{\rm u}=100\mev$, which is sufficiently high to prevent strange quark nuggets with small baryon number $A<10^{52}$ from surviving.
The weak-interaction equilibrium of up, down and strange flavours in strange quark matter is independent of the process by which quarks assemble into nucleons and therefore does not suppress evaporation~\cite{alcock_evaporation_1985}. 
We therefore propose a novel mechanism: if strange quarks undergo a phase transition into strangeons, the evaporation process can be strongly suppressed. This suppression arises because the evaporation of strangeon nuggets requires the 
weak-interaction conversion from strangeons to nucleons on the surface of strangeon nuggets, which is $<10^{-13}$ times the rate of strong-interaction quark assembling when strange quark matter evaporates nucleons~\cite{xu_solution_2014,wang_optical_2017,li_polar_2026}.
Within this framework, the next crucial step is to determine the temperature $T$ of the transition as the Universe cools down, such that strangeons are formed before significant evaporation of strange quarks takes place.

In this work, we determine the stable phase by requiring thermodynamic stability~\cite{callen_thermodynamics_1985}. Specifically, at a given temperature $T$, the phase with the higher pressure $P$ at a fixed baryon chemical potential $\mu_{\rm B}$ is energetically favored. Consequently, pressure-versus-baryon-chemical-potential $P(\mu_{\rm B})$ relations of the two phases must intersect at least once, and the intersection point defines the first-order transition chemical potential. In the left panel of Fig.~\ref{fig:pmuBTmuB}, we show the pressure $P$ as a function of the baryon chemical potential $\mu_{\rm B}$ for strange quark matter and strangeon matter at different temperatures. We focus on the case with a single intersection, in which strangeon matter is the stable phase at low baryon densities, while strange quark matter is stable at high densities. In Fig.~\ref{fig:pmuBTmuB}, the results show that, at $T=50\mev$, the transition occurs at $\mu_B = 863~\mathrm{MeV}$. As the temperature decreases, the transition point shifts to higher values of $\mu_{\rm B}$, reaching $\mu_B > 1100\mev$ at lower temperatures. This suggests that, as the Universe cools down, strangeon matter is expected to be the favored state.

Within the scenario introduced in Sec.~\ref{sec:formalism}, we determine the phase-transition chemical potentials at different temperatures in the $T–\mu_{\rm B}$ plane, as shown in the right panel of Fig.~\ref{fig:pmuBTmuB}. The red curve denotes the first-order transition line. Its solid segment corresponds to transition points occurring at positive pressure, whereas the dashed segment indicates regions where the pressure is negative and thus unphysical for the early Universe. The red dot in Fig.~\ref{fig:pmuBTmuB} marks the strange-quark–strangeon phase-transition point at $p=0$. The results show that strange quark matter is the thermodynamically favored phase at high temperatures and high baryon chemical potentials, while strangeon matter is favored at low temperatures and low baryon chemical potentials. As the temperature increases, both the baryon chemical potential and the pressure at the phase-transition point decrease. Over the temperature range in which the strange-quark–strangeon phase transition can occur, the pressure of the Universe remains small.
Our results indicate that $p = 0.81~\mathrm{MeV/fm^3}$ at $T = 50\mev$. 
For the parameter set used in Fig.~\ref{fig:pmuBTmuB}, this corresponds to a deviation of approximately $0.4\mev$ from the phase-transition temperature at zero pressure, which is negligible. In the following calculations, we therefore determine the phase-transition temperature by the value indicated by the red point in Fig.~\ref{fig:pmuBTmuB}.

    We propose during the first-order QCD phase transition in the early Universe, the temperature is sufficiently high to form strange quark nuggets initially.
    Their evolution is described by Eqs.~(\ref{eq:dAdt}) and~(\ref{eq:detT_nug}) (see Appendix~\ref{appendix: evap} for detailed derivation). As the temperature decreases, strangeon matter becomes more stable, potentially triggering a transition from strange quark nuggets into strangeon nuggets, causing a significant reduction of nugget evaporation rate, to approximately $10^{-13}$ times that of the strange quark nuggets~\cite{xu_solution_2014,wang_optical_2017,li_polar_2026}, rendering it negligible\footnote{If we use a evaporation rate which is $10^{-13}$ times that of the strange quark nuggets in Eq.~(\ref{eq:dAdt}), the evaporation is significant ($A/(dA/dt)<10^{-6}\,\rm s$) when $A<10^{23}$. This baryon number is much smaller than that in the particle horizon at cosmic QCD phase transition ($10^{49}$), so it does not affect the conclusion.}. If such a transition occurs after the strange quark nuggets lose 15\% of their baryons, the ratio of baryonic matter to strangeon matter, is about $\overline{\varepsilon}_{\rm baryon}:\overline{\varepsilon}_{\rm strangeon}=15\%:85\%$, consistent with the ratio of baryonic matter to dark matter~\cite{aghanim_planck_2020}. Consequently, these relic strangeon nuggets could represent a viable candidate for DM.

    We investigate the evolution of strange quark nuggets across a range of initial baryon numbers $A_0$ from $10^7$ to $10^{57}$. This range of $A_0$ is chosen to ensure that the strange quark nuggets can be treated as bulk matter and are unlikely to collapse into a black hole. Regarding the binding energy per baryon $\delta E$, we set its range from 0 to $75\mev$ (in MIT bag model, it corresponds from $B^{1/4}=148\mev$ to $159\mev$) to align with the Bodmer-Witten hypothesis. The initial temperature at which strange quark nuggets form is $T_{\rm u0}=100\,{\rm MeV}$, corresponding to the QCD phase transition temperature. Fig.~\ref{fig:T15} shows the temperature of the strange quark nuggets when they lose 15\% of their baryons. 
    The evaporation rate depends on the equilibrium between the cooling and heating of the strange quark nuggets. According to Eqs.~(\ref{eq: L_nu})-(\ref{eq:detT_nug}), cooling occurs as a result of energy loss caused by evaporating nucleons, while heating is due to the neutrino-quark scattering.
    The evaporation rate decreases exponentially with increasing $\delta E$. As a result, a higher $\delta E$ value slows down the cooling process, leading to a higher $T_{\rm nug,\,15}$.
    On the other hand, nuggets with $A_0>10^{40}$ have radius larger than the mean free path of the neutrinos, which significantly enhances neutrino-quark scattering and results in more effective heating from neutrinos. Consequently, the evaporation rate increases, reducing the time required to evaporate 15\% of baryons and leading to a higher $T_{\rm nug,\,15}$.

    To find the possibility that survived strangeon nuggets can contain 85\% baryonic matter in the Universe, for each strange quark nugget with a certain $\delta E$ and $A_0$, we further explore the feasible parameter space of $m_1$, $m_2$, $u_0^{(0)}$, $r_0^{(0)}$ of the strangeon matter that allow a strangeon nugget is more stable than a strange quark nugget at $T_{\rm nug,15}$, then summarize all the parameter spaces. 
    The left panel of Fig.~\ref{fig:params} shows the summarized parameter space with red area is allowed.
    The allowed region for $m_1$ and $m_2$ is primarily determined by the condition $m_1>m_2$. Meanwhile $u_0^{(0)}$ exhibits a tendency toward larger values, whereas the allowed range for $r_0^{(0)}$ shows the opposite trend.
    This behavior arises because an increase in $u_0^{(0)}$ represents a deeper potential well and a stronger repulsive force between strangeons when they approach each other. This results in a stiffer EoS. The formula for $u_0^{(0)}$ changing with temperature determines that the larger $u_0^{(0)}$, the more sensitive the EoS is to temperature.
    This will elevate the phase transition pressure on $p$-$\mu_{\rm B}$ curves, and consequently lead to a higher transition temperature from strange quark nuggets to strangeon nuggets, allows strangeon nuggets to survive more easily.
    Meanwhile a decrease in $r_0^{(0)}$ corresponds to stronger attraction between strangeons, making the EoS softer to counteract the effect of large $u_0^{(0)}$. The $m_1$ and $m_2$ fall within the range that can adapt the force range to $\sim r_0^{(0)}$.

    Strangeon matter can not only survive as DM in the early Universe, it can also exist in pulsars and account for current static astronomical observations. By self-consistently reducing the phenomenological model of Sec.~\ref{subsec:Strangeon} to zero temperature, we predict the global properties of strangeon stars. Here, 
    we calculate the mass-radius relationship of strangeon stars using the Tolman-Oppenheimer-Volkoff (TOV) equations \cite{oppenheimer_massive_1939}. Additionally, the tidal deformability $\Lambda$-mass curves for strangeon stars is derived through the gauge perturbation equations \cite{hinderer_tidal_2008}. The $M$~-~$R$ and $\Lambda$~-~$M$ properties of several sets of parameters are shown in Appendix \ref{appendix: star}. 
    We conducted a Bayesian analysis to investigate the parameters $m_1$, $m_2$, $u_0^{(0)}$, and $r_0^{(0)}$ of the strangeon EoS.
    The posteriror distribution $\mathcal{P}(\theta|D)$ is obtained from the prior distribution $\pi(\theta)$ and the likelihood function $\mathcal{L}(D|\theta)$ in Bayes' theorem,
    \begin{equation}
        \begin{aligned}
            \mathcal{P}(\theta|D)\propto\pi(\theta)\mathcal{L}(D|\theta)~,
        \end{aligned}
    \end{equation}
    where $\theta$ represents all the inferenced parameters, $D$ is the used observational dataset.
    The prior distribution of four srangeon EoS parameters is introduced in Sec.~\ref{subsec:Strangeon}, i.e. $m_1\sim U(60\,{\rm MeV},1200\,{\rm MeV})$, $m_2\sim U(60\,{\rm MeV},1200\,{\rm MeV})$, $u_0^{(0)}\sim U(0,400\,{\rm MeV})$, $r_0^{(0)}\sim U(1.5\,{\rm fm},4.5\,{\rm fm})$, where $U(\cdot)$ denotes uniform distribution. The likelihood function contains three non-observational terms: one is $m_1>m_2$ which indicates the repulsive interaction has a shorter range; one is the requirement that the density of strangeon matter under zero pressure exceeds $n_0$; one is the constraints from the left panel of Fig.~\ref{fig:params} to ensure the inferred posterior parameter space is a subspace of the region that strangeon nuggets could survive as DM. The datasets considered in the likelihood function contain the mass-radius measurements by NICER X-ray data from PSR J0030+0451~\cite{vinciguerra_updated_2024,riley_nicer_2019,miller_psr_2019} ($M=1.34^{+0.15}_{-0.16}M_\odot, R=12.71^{+1.14}_{-1.19}\,{\rm km}$), PSR J0437-4715~\cite{choudhury_nicer_2024,rutherford_constraining_2024}($M=1.418\pm 0.037M_\odot, R=11.36^{+0.95}_{-0.63}\,{\rm km}$), and the heavy pulsar PSR J0740+6620~\cite{salmi_radius_2024,miller_radius_2021,riley_nicer_2021}($M=2.072^{+0.067}_{-0.066}M_\odot, R=12.39^{+1.30}_{-0.98}\,{\rm km}$). Additionally, the tidal deformability $\Lambda_{1.4}$ data from the LIGO/Virgo gravitational wave event GW170817~\cite{ligo_observation_2017,ligo_gw170817_2018,abbott_properties_2019} is used in the likelihood. 
    The results are shown in the right panel of Fig.~\ref{fig:params}. The contour here represent the confidence levels of 50\% for the dark blue and 90\% for the light blue, respectively. 
    The preferred parameter estimates are: $m_1=710^{+312}_{-263}\,\mev$, $m_2=239^{+98}_{-90}\,\mev$, $u_0^{(0)}=303^{+71}_{-110}\,\mev$, $r_0^{(0)}=2.06^{+0.36}_{-0.18}\,{\rm fm}$.

    \section{Conclusions}\label{sec:summary}

    Bodmer-Witten hypothesis suggests the possibility to form strange quark nugget during the first order QCD phase transition of the early Universe. However, due to the rapid evaporation, they could not survive with small baryon number.
    In this work, we propose a further phase transition from strange quark nuggets to strangeon nuggets. 
    The weak-interaction conversion from strangeons to nucleons on the surface of strangeon nuggets is approximately $10^{-13}$ times slower than the strong-interaction quark assembling when strange quark nugget evaporates nucleons. 
    Therefore, after this phase transition, the evaporation stopped, and the relic strangeon nuggets could serve as DM in the early Universe.
    
    We applied MIT bag model to the strange quark matter. 
    As for the strangeon matter, we introduce the temperature effect to the double Yukawa potential to describe the strong interaction and the EoS of it. 
    The phase diagram in Fig.~\ref{fig:pmuBTmuB} shows that strange quark matter is more stable at high temperature and high chemical potential, while strangeon matter is more stable at low temperature and low chemical potential. When the strange quark nuggets cool down, they may undergo a phase transition to strangeon nuggets.

    We calculate the evaporation of strange quark nugget with initial mass number $A_0$ from $10^7$ to $10^{57}$ and binding energy $\delta E$ in the range 0 to $75\mev$. Those strange quark nuggets can lose 15\% of their baryons at temperatures approximately between 40\,MeV and 70\,MeV. The remaining 85\% baryon mass in strangeon nuggets indicate that it could serve as DM. We further explore the possible parameter space that allow the phase transition from strange quark matter to strangeon matter at the appropriate temperature to evaporate 15\% of the baryons from strange quark nuggets. 
    The parameter ranges are about $400\mev < m_1 < 1200\mev$, $60\mev < m_2 < 700\mev$, $120\mev < u_0^{(0)} < 400\mev$, and $1.5\,{\rm fm} < r_0^{(0)} < 4.1\,{\rm fm}$, shown in the left panel of Fig.~\ref{fig:params}. A Bayesian inference is further applied to ensure that under the zero temperature limit, the calculated $M$~-~$R$ relations and the tidal deformability of the pulsar-like compact stars are consistent with observations. The inferred parameters are $m_1=710^{+312}_{-263}\,\mev$, $m_2=239^{+98}_{-90}\,\mev$, $u_0^{(0)}=303^{+71}_{-110}\,\mev$, $r_0^{(0)}=2.06^{+0.36}_{-0.18}\,{\rm fm}$. 

    There are still some problems in this work that can be further analyzed. The temperature of the cosmic QCD phase transition, also the initial temperature of the strange quark nuggets, is fixed to $100\mev$, but there is uncertainty about this temperature. A higher initial temperature will cause evaporation more quickly, thereby narrowing the range of strangeon matter EoS parameters that can satisfy the dark matter ratio. Our rough calculations indicate that the allowed parameter space will not change too much when the initial temperature varying between $80\mev$ and $120\mev$.  Besides, we ignore the evaporation of strangeon nuggets. More precise calculations can consider the effects of evaporation of these extremely small strangon nuggets, such as the dark matter mass spectrum and its impact on subsequent galaxy formation and evolution.
    In future, the strong nuggets can be searched through terrestrial acoustic measurements~\cite{qi_detect_2025}. Exploring the nature of DM and strong nuggets will further enhance our understanding of the early Universe and the QCD phase diagram.

    \medskip
	\acknowledgments
	We greatly thank Dr. Yu-Cheng Qiu for his inspiring and insightful discussions. We also thank Professor Kejia Lee, Professor Fangzhou Jiang, and Dr. Zhiqiang Miao for their valuable discussions and comments. We thank PKU pulsar group for the discussions. This work is supported by the National SKA Program of China (Grant. No. 2020SKA0120100).
	W.-L. Y. is supported by the China Postdoctoral Science Foundation (Grant. No. 2025M773418). Yudong Luo is supported by the National Natural Science Foundation of China (NSFC, No. 12335009), the China Postdoctoral Science Foundation (Grant. No. 2025T180924) and the Boya Fellowship of Peking University.  
    We conduct the Bayesian analysis using code \textit{emcee}~\cite{foremanmackey_emcee_2013}.

    \onecolumngrid

	\appendix
    \section{Thermodynamic derivations of the strangeon matter system at finite temperature}\label{appendix: thermo}

	The definitions of thermodynamic quantities in this paper are as follows,
	\begin{equation}\label{eq:definition}
		\begin{aligned}
			\Omega = -p~,\qquad
			\varepsilon = U/V~,\qquad
			s = S/V~,\qquad
			n = N/V~, 
		\end{aligned}
	\end{equation}
	in which $U$ is the total internal energy, $p$ is the pressure, $V$ is the total volume, $S$ is the total entropy, $N$ is the total number of particles, $\varepsilon$ is the total internal energy density, $s$ is the entropy density, $n$ is the particle number density and $\Omega$ is the grand canonical potential density.
	The relation between the thermodynamic quantities reads
	\begin{equation}\label{eq:thermalrelation}
		\begin{aligned}
			& U &= -pV+TS+\mu N~, \\
		\end{aligned}
	\end{equation}
	where $T$ is the temperature, $\mu$ is the chemical potential per particle.
	According to Eq.~(\ref{eq:thermalrelation}), Eq.~(\ref{eq:definition}) can be rewritten as
	\begin{equation}
		p+\varepsilon = Ts+\mu n~.
	\end{equation}
	The differential form of the first law of thermodynamics reads
	\begin{equation}\label{eq:firstlaw}
		dU = -pdV+TdS+\mu dN~.
	\end{equation}
	The following differential relations can be obtained from Eq.~(\ref{eq:definition}) to Eq.~(\ref{eq:firstlaw}),
	\begin{equation}\label{eq:difffirstlaw}
		\begin{aligned}
			&d(\Omega V) = -pdV-SdT-Nd\mu~, 
			&&d\Omega = \frac{d(\Omega V)}{V} - \frac{(\Omega V)dV}{V^2} 
			= -\frac{pdV}{V}-sdT-nd\mu+\frac{pdV}{V}
			= -sdT-nd\mu~, \\
			&dp = sdT+nd\mu~,
			&&d\varepsilon = Tds+\mu dn~.
		\end{aligned}
	\end{equation}

	From the properties of total differentiation, it can be concluded that $n=(\partial p/\partial \mu)|_T$, $\mu=(\partial \varepsilon /\partial n)|_s$. 
	
	Because the input quantities in the strangeon EoS are the number density $n$ and the temperature $T$, we calculate the pressure $p$ using the following formula. When the total number of particles in the system $N$ is conserved, we have
	\begin{equation}\label{eq:pressure2}
		\begin{aligned}
			p = \frac{1}{dV}(TdS-dU) 
			= T \frac{d(sV)}{dV}-\frac{d(\varepsilon V)}{dV} 
			= T \frac{d(sN/n)}{d(N/n)}-\frac{d(\varepsilon N/n)}{d(N/n)} 
			= -T n^2 \frac{d(s/n)}{dn} + n^2 \frac{d(\varepsilon/n)}{dn}~.
		\end{aligned}
	\end{equation}
	For a given temperature $T$, the pressure $p$ is obtained by calculating the differential on the number density $n$ of $\varepsilon/n$ and $s/n$, then substitute into Eq.~(\ref{eq:pressure2}).

	The energy density $\varepsilon$ can be calculated by summing up the potential energy, the rest energy, and the thermal kinetic energy,
    \begin{equation}
    \begin{aligned}
		\varepsilon = \varepsilon_{\rm p} + m n + \varepsilon_{\rm k}\ .
        \end{aligned}
	\end{equation}
    
    The entropy dendity $s$ is obtained from the probability of occupation $q$, $0\leq q\leq 1$:
	\begin{equation}
		s = -\sum_{d}[q(d)\ln q(d)+(1-q(d))\ln (1-q(d))]~,
	\end{equation}
	in which $d$ represents all quantum states and degeneracy.
	Strangeons are non-relativistic due to their large rest mass, they have small quantum wave packets, so we consider them as classical particles. For a given temperature, strangeons follow the Maxwell-Boltzmann distribution, their probability of occupation $q$ is given by
	\begin{equation}
		q(k)=(2\pi)^3n\left(\frac{1}{2\pi m T}\right)^{3/2}\exp \left(-\frac{k^2}{2mT}\right)~,
	\end{equation}
	in which $k$ represents the modulus of the momentum of one particle $\boldsymbol{k}$.
	When $T>2$\,MeV, $q(k)\leq q(k=0)<0.1$, it fits the classic statistical limits, $q(k)\ll 1$.
	The entropy density reads

	\begin{equation}\label{eq:MBentropy}
		\begin{split}
			s &= -\int_0^\infty \frac{4\pi k^2}{(2\pi)^3} [q(k)\ln q(k)+(1-q(k))\ln (1-q(k))] dk 
			\simeq -\int_0^\infty \frac{4\pi k^2}{(2\pi)^3} [q(k)\ln q(k)-q(k)] dk \\
			&= n\left[1-\int_0^\infty \frac{4\pi k^2}{(2\pi)^3}
			\left(\frac{2\pi}{m T}\right)^{3/2}
			\exp \left(-\frac{k^2}{2mT}\right)\left(\ln n
			+\frac{3}{2}\ln \frac{2\pi}{mT}
			-\frac{k^2}{2mT}\right) dk \right] \\
			&= \frac{5}{2} n
			+n\ln\left[\left(\frac{mT}{2\pi}\right)^{3/2}n^{-1}\right]~.
		\end{split}
	\end{equation}

	In the case of Maxwell-Boltzmann distribution, the thermal kinetic energy density $\varepsilon_{\rm k}=3nT/2$, and $\frac{d(\varepsilon_{\rm k}/n)}{dn}=0$. The temperature-correction of the pressure $p$ is the first term in Eq.~(\ref{eq:pressure2}), $-Tn^2\frac{d(s/n)}{dn}=nT$.
	The chemical potential $\mu$ calculated from $\varepsilon=3nT/2$, $p=nT$, and Eq.~(\ref{eq:MBentropy}) reads
	\begin{equation}
		\mu = \frac{\varepsilon+p-Ts}{n}=T\ln n+\frac{3}{2}T\ln \left(\frac{2\pi}{mT}\right)~.
	\end{equation}

\section{The formalism for the evaporation of strange quark nuggets}\label{appendix: evap}
Consider the chemical equilibrium of a nugget with baryon number $A$ absorbing a 
    nucleon and a nugget with baryon number $(A+1)$ evaporating a 
    nucleon, i.e., $(A+1)\leftrightarrow (A)+{\rm (nucl)}$.
   The absorption rate per unit volume, $\nu_{\rm abs}$, can be written as~\cite{alcock_evaporation_1985}
\begin{equation}
		\begin{aligned}
			\nu_{\rm abs} = n_{\rm nug} n_{\rm nucl} \sigma_{\rm abs} \overline{v_{\rm nucl}}~,
		\end{aligned}
	\end{equation}
where $n_{\rm nug}$ denotes the number density of strange quark nuggets, $n_{\rm nucl}$ is the number density of ambient nucleons, $\sigma_{\rm abs}$ presents the effective absorption cross section, and
$\overline{v_{\rm nucl}}=\sqrt{T/(2\pi m_{\rm nucl})}$ 
is the mean nucleon velocity.
The absorption cross section is taken to be proportional to the geometric cross section,
$\sigma_{{\rm abs},i} = f_i~4\pi r_{\rm nug}^2,$
where $0<f_i<1 $ is the absorption probability per collision for particle species $i$. The value of $f_i $ depends on the type of absorbed particle. The relation between the nugget-radius and the baryon number $A$ is $(4\pi/3) r_{\rm nug}^3n_{\rm B}=A$.

    The evaporation rate is proportional to the nugget-number-density
	\begin{equation}\label{eq:evap_rate}
		\begin{aligned}
			\nu_{\rm evap} = \kappa n_{\rm nug}~.
		\end{aligned}
	\end{equation}

	  Reaction equilibrium drives chemical potentials obey
    \begin{equation}
		\begin{aligned}
			\mu_{{\rm gas}(A+1)} = \mu_{{\rm gas}(A)} + \mu_{{\rm gas(nucl)}} + \delta E~,
		\end{aligned} \label{eq: chemical eqli}
	\end{equation}with $\delta E$ representing the binding energy.
    Since the working temperature is much lower than the neutron mass, the relevant particles in the idea gas are nonrelativistic and nondegenerate. The chemical potential of each particle species $i$ is therefore given by
	\begin{equation}
		\begin{aligned}
			\mu_{{\rm gas},i} = T \ln \left[\left(\frac{2\pi}{m_i T}\right)^{3/2}\frac{n_i}{g_i}\right]~.
		\end{aligned} \label{eq: particle chemical P}
	\end{equation}
	
    Here, $m_i$, $n_i$, and $g_i$ denote the particle mass, number density, and internal degeneracy, respectively.

    For large nuggets with $ A \gg 1 $, the degeneracy factor satisfies $g_{A+1}/g_A \to 1$. For nucleons, $g_{\rm (nucl)}=2$ due to spin. Combining Eqs.~(\ref{eq: chemical eqli}) and (\ref{eq: particle chemical P}) implies:
	\begin{equation}
		\begin{aligned}
			\frac{n_{\rm (nucl)}n_{(A)}}{n_{(A+1)}} = 2\exp\left(-\frac{\delta E}{T}\right) \left(\frac{m_{\rm (nucl)} T}{2\pi}\right)^{3/2}~.
		\end{aligned} \label{eq: nnbnA/nA+1}
	\end{equation}
    Imposing Eq.~(\ref{eq: nnbnA/nA+1}), together with the thermal equilibrium, in which strange quark nuggets absorb and emit nucleons at equal rates, $\nu_{{\rm evap},(A+1)}=\nu_{{\rm abs},(A)}$, the evaporation rate $\kappa$ can be derived as
	
	\begin{equation}
		\begin{aligned}
			\kappa=\frac{m_{\rm (nucl)}T^2}{2\pi^2}\exp\left(-\frac{\delta E}{T}\right)\sigma_{\rm abs}~.
		\end{aligned}
	\end{equation}
	The evaporation process removes internal thermal energy from the nugget. The high density of strange quark matter results in a large thermal conductivity. Thus, it is acceptable to treat the strange quark nugget as having a uniform temperature.
	
    Each evaporated nucleon carries away an energy of $\delta E+2T_{\rm nug}$ energy, causing the nugget temperature $T_{\rm nug}$ to fall below the temperature of the Universe $T_{\rm u}$.
    
    When $T_{\rm nug}<\delta E$, 
    neutrinos in the Universe can provide the sustained external heat source required for continued evaporation. Consider the absorption of neutrinos with a mean free path of $l_{\nu}=G_{\rm F}^{-2}\mu_{q}^{-2}T^{-3}$, where $\mu_{q}\simeq 300\,{\rm MeV}$ is the quark chemical potential inside 
    the nugget~\cite{alcock_evaporation_1985}, the net heating rate is
	\begin{equation}\label{eq: L_nu}
		\begin{aligned}
			L_{\nu} = 4\pi r_{\rm nug}^2 \left(\frac{7\pi^2}{160}\right)[T_{\rm u}^4p(r_{\rm nug},T_{\rm u}) - T_{\rm nug}^4p(r_{\rm nug},T_{\rm nug})]~,
		\end{aligned}
	\end{equation}
	in which
	\begin{equation}
		p(r,T) = \left\{
		\begin{aligned}&~1,~r_{\rm nug}>3l_{\nu}/4~, \\
			&~\frac{4r_{\rm nug}}{3l_{\nu}},~r_{\rm nug}<3l_{\nu}/4~,
		\end{aligned}
		\right.
	\end{equation}
	is the probability of neutrino absorption during a single neutrino-nugget interaction.

    Evaporation leads to a higher particle density near the nugget surface than in the distant region. Far away from the nuggets, the pressure generated by electrons, positrons, and photons is $p_{\rm far}=(11\pi^2/180)T_{\rm u}^4$. Near the nuggets, the pressure receives an additional contribution from nucleons and is given by $p_{\rm near}=(11\pi^2/180)T_{\rm nug}^4+\Sigma_{i=n,p}~n_i T_{\rm nug}$. The pressures are equal near and far from the strange quark nuggets. Owing to the small Coulomb barrier of the nugget ($\sim 10 \mev$) and the relatively high temperature, the effect of the Coulomb barrier can be neglected. Consequently, the evaporation rates of protons and neutrons are nearly identical:
    \begin{equation}
        \begin{aligned}
            n_p=n_n=\frac{1}{2}\frac{11\pi^2}{180}\left(T_{\rm u}^4-T_{\rm nug}^4\right)~.
        \end{aligned}
    \end{equation}
   
    These conditions lead to the rate of baryon-number loss of a single nugget,
	\begin{equation}\label{eq:dAdt}
		\begin{aligned}
			\frac{{\rm d} A}{{\rm d} t} = \left[\frac{m_n T_{\rm nug}^2}{2\pi^2}\exp{\left(-\frac{\delta E}{T_{\rm nug}}\right)}\right. 
			 \left. - \frac{11\pi^2}{360}(T_{\rm u}^4-T_{\rm nug}^4)\left(\frac{T_{\rm nug}}{2\pi m_n}\right)\right](\sigma_{{\rm abs},n}+\sigma_{{\rm abs},p})~.
		\end{aligned}
	\end{equation}
	
    The internal thermal energy of the nuggets should balance energy gain and loss. This requirement determines the nugget temperature $T_{\rm nug}$ through
	\begin{equation}\label{eq:detT_nug}
		\begin{aligned}
		    L_\nu + \frac{{\rm d} A}{{\rm d} t} (\delta E+2T_{\rm nug}) = 0~.
		\end{aligned}
	\end{equation}
	
	With these relations, the time evolution of the baryon number and temperature of the nuggets, as well as the temperature evolution of the Universe, can be calculated.

\section{Properties of strangeon star}\label{appendix: star}

    By integrating the Tolman-Oppenheimer-Volkoff (TOV) equations \cite{oppenheimer_massive_1939} and the the gauge perturbation equations \cite{hinderer_tidal_2008}, the mass-radius and $\Lambda$-mass relationships of strangeon stars are calculated. Fig.~\ref{fig:pe,mr,lm} shows the $M$~-~$R$ plot and the $\Lambda$~-~$M$ plot of some representative strangeon matter EoS parameters. A stiffer EoS will result in a larger mass at the same radius or $\Lambda$. The selection of five sets of parameters can display the individual effects of each parameter on the properties of strangeon stars. An increase in $u_0^{(0)}$ represents a deeper potential well and a stronger repulsive force between strangeons when they approach each other, resulting in a stiffer EoS. A decrease in $r_0^{(0)}$ corresponds to stronger attraction between strangeons, making the EoS softer. The $m_1$ and $m_2$ fall within the range that can adapt the force range to $\sim r_0^{(0)}$. Because the force range of repulsive interaction is shorter, $m_1>m_2$. A larger $m_1$ makes the repulsive effect stronger and the EoS stiffer, while $m_2$ is the opposite, representing the attractive effect. 
    \begin{figure*}
		
        \includegraphics[width=0.98\textwidth]{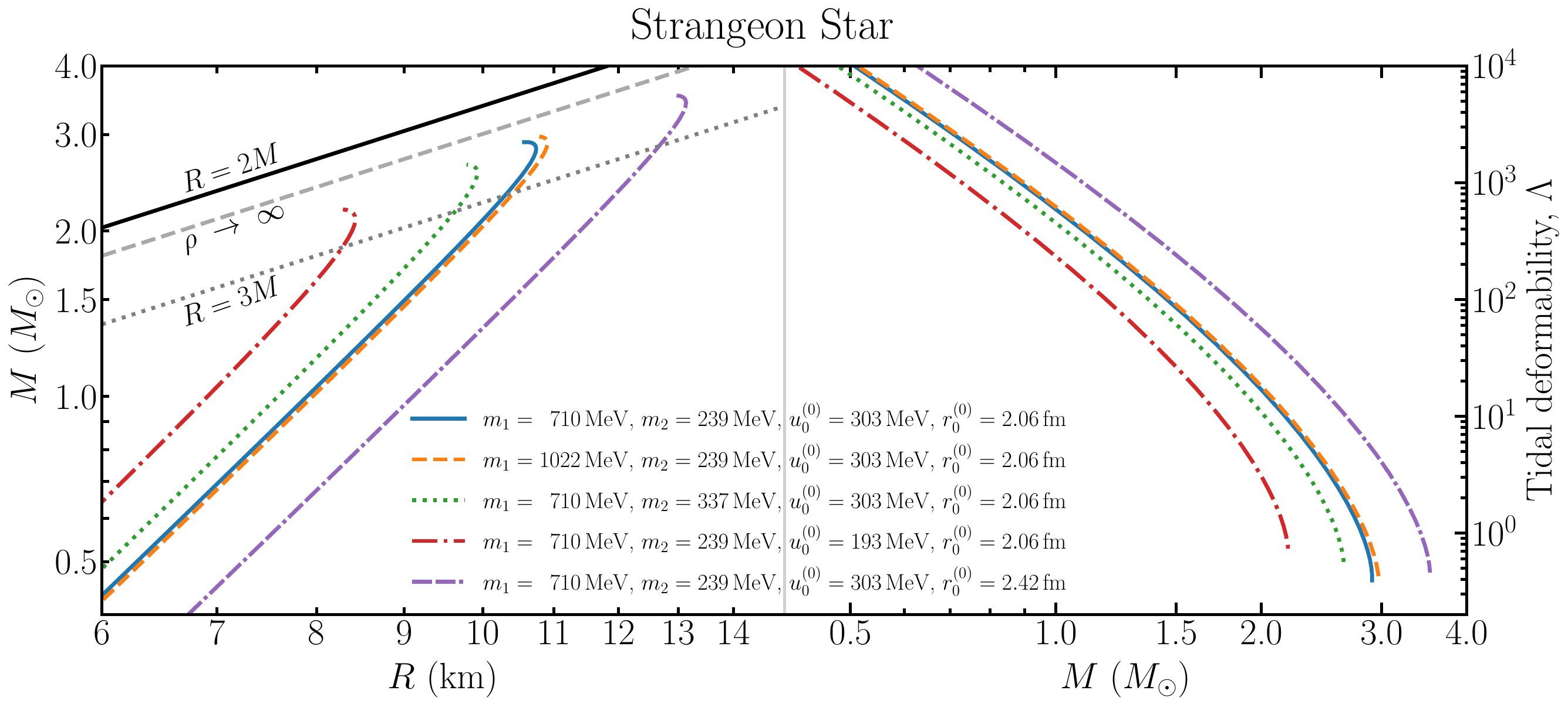}
        
		\caption{
        The left panel is the mass-radius curves of the strangeon stars calculated through the TOV equations~\cite{oppenheimer_massive_1939}. The right panel is the tidal deformability-mass curves of the strangeon star calculated by the gauge perturbation equations~\cite{hinderer_tidal_2008}. The solid-blue line shows that four EoS parameters are equal to their average value inferred in the right panel of Fig.~\ref{fig:params}, i.e. $m_1=710\,\mev$, $m_2=239\,\mev$, $u_0^{(0)}=303\,\mev$, $r_0^{(0)}=2.06\,{\rm fm}$. The other four lines each change one of the four parameters. The dashed-orange line represents the change in $m_1$ to $1022\,\mev$. The dotted-green line alters $m_2$ to $337\,\mev$. The dash-dotted-red line turns $u_0^{(0)}$ to $193\,\mev$. The dash-dash-dotted-purple line changes $r_0^{(0)}$ to $2.42\,{\rm fm}$.}\label{fig:pe,mr,lm}
	\end{figure*}

\twocolumngrid

\end{document}